# Competition of electronic correlation and reconstruction in La$_{1-x}$Sr$_x$TiO$_3$/SrTiO$_3$ heterostructures

Xueyan Wang[1†*], Lin Sun[2†], Chen Ye[1], Zhen Huang[3], Kun Han[3], Ke Huang[1], Allen Jian Yang[1], Shengwei Zeng[2], Xian Jun Loh[4,5], Qiang Zhu[1,4,5], T. Venkatesan[6], Ariando Ariando[2*], X. Renshaw Wang[1,7*]

[1]School of Physical and Mathematical Sciences, Nanyang Technological University, Singapore 637371

[2]Department of Physics & NUSNNI-Nanocore, National University of Singapore, Singapore 117411

[3]Institutes of Physical Science and Information Technology, Anhui University, Hefei, Anhui, China 230601

[4]Institute of Materials Research and Engineering (IMRE), A*STAR, Singpore 138634

[5]Institute of Sustainability for Chemicals, Energy and Environment (ISCE2), Agency for Science, Technology and Research (A*STAR), 1 Pesek Road, Jurong Island, Singapore, 627833

[6]Center for Quantum Research and Technology, University of Oklahoma, Norman OK 73019

[7]School of Electrical and Electronic Engineering, Nanyang Technological University, Singapore 639798

†These authors contributed equally to this work

*Email: xueyan.wang@ntu.edu.sg;

phyarian@nus.edu.sg;

renshaw@ntu.edu.sg

# Abstract

Electronic correlation and reconstruction are two important factors that play a critical role in shaping the magnetic and electronic properties of correlated low-dimensional systems. Here, we report a competition between the electronic correlation and structural reconstruction in $La_{1-x}Sr_xTiO_3/SrTiO_3$ heterostructures by modulating material polarity and interfacial strain, respectively. The heterostructures exhibit a critical thickness ($t_c$) at which a metal-to-insulator transition (MIT) abruptly occurs at certain thickness, accompanied by the coexistence of two- and three-dimensional (2D and 3D) carriers. Intriguingly, the $t_c$ exhibits a V-shaped dependence on the doping concentration of Sr, with the smallest $t_c$ value at $x = 0.5$. We attribute this V-shaped dependence to the competition between the electronic reconstruction (modulated by the polarity) and the electronic correlation (modulated by strain), which are borne out by the experimental results, including strain-dependent electronic properties and the evolution of 2D and 3D carriers. Our findings underscore the significance of the interplay between electronic reconstruction and correlation in the realization and utilization of emergent electronic functionalities in low-dimensional correlated systems.

# Main text

Oxide interfaces have recently garnered significant research attention due to their intriguing properties and functionalities that differ from those of their bulk counterparts[1-4]. Various interesting physical phenomena, such as two-dimensional electron gas (2DEG) at the interface between insulating layers[5], superconductivity[6], and ferromagnetism[7], have been observed. In particular, the interfacial 2DEG is widely believed to originate from the electronic reconstruction, *i.e.* charge transfer forced by the polar catastrophe at the polar/non-polar interface[8]. By designing the polar overlayer[9] or the surface adsorption[10], the strength of the internal electric field in the polar material can be tuned, resulting in a dramatic change of the electronic properties, such as Rashba spin-orbit coupling (SOC)[11-14] which interacts with the electronic Coulomb interaction. Moreover, oxide interfaces feature besides electronic reconstruction a strong electron-electron interaction. Hence, the Coulomb interaction from electronic correlation[15], depending on the interfacial strain[16] and the type of the rare-earth cations[17], could compete against the Rashba SOC from electronic reconstruction, leading to the emergence of unconventional electronic phases[18-21]. Intensive investigations have been conducted to study each mechanism individually[15,22-24]. However, electronic correlation and reconstruction coexist at certain system and these properties jointly govern the electrical characteristics of these oxide systems. To investigate the primary driving mechanism of the rich physical phenomena, a deeper understanding of the interplay between electronic correlation and reconstruction in an ultrathin complex oxide system is necessary. Given its complexity and indivisibility, this poses a significant challenge.

A doped interfacial system serves as an ideal platform for investigating the effect of the interplay on electronic properties[25]. Sr-doped LaTiO$_3$ (La$_{1-x}$Sr$_x$TiO$_3$, LSTO) has garnered significant attention due to its potential for a variety of applications, such as solid oxide fuel cells[26], transparent conductors[27,28], and buffer layer for high-temperature superconductors[29]. Doping Sr into LaTiO$_3$ (LTO) thin films leads to the emergence of a conducting phase due to the band filling with each Sr$^{2+}$ replacing La$^{3+}$. The doping provides an additional electron to the Ti site, facilitating the transition from the 3$d^1$ to the 3$d^0$ configuration. More importantly, doping of Sr also introduces a modulation of both electronic correlation and reconstruction simultaneously.

The strength of electronic correlation can be estimated by the 3$d$ $t_{2g}$ electronic orbital bandwidth[30]. Structurally, Sr doping transforms the LSTO from *Ibmm* (0.2 ≤ $x$ ≤ 0.7) to *Pbnm* (0.7 ≤ $x$ ≤ 1)[28] with oxygen octahedral rotation that leads to a change in the Ti-3$d$ bandwidth. Also, the interfacial strain induced by the lattice mismatch between LSTO and STO will decrease as the Sr concentration is varied from $x$ = 0 (LTO, 3.97 Å) to $x$ = 1 (STO, 3.905 Å), modifying the LSTO band structure accordingly[31]. Concurrently, because the LSTO transforms from a polar La$^{3+}$Ti$^{3+}$O$_3$ configuration for $x$ = 0 to a non-polar Sr$^{2+}$Ti$^{4+}$O$_3$ configuration for $x$ = 1, the effect of polarization catastrophe-induced electronic reconstruction is expected[32]. Thus, a systematic study on the electronic properties of the LSTO/STO heterostructure will allow us to probe the interplay between the two phenomena, *i.e.* the electronic reconstruction and correlation, and to prove usefulness in optimizing the electronic properties in complex oxide systems for further applications.

In this study, we investigated the evolution of the electronic properties of the LSTO/STO heterostructures by varying the Sr-dopant concentration, the LSTO thickness, and the substrate-induced interfacial (biaxial) strain. With the Sr-doping ranging from $x$ = 0.1 to 0.9 and the LSTO thickness ranging from a few to up to 300 unit cells (uc), the effect of cation substitution, biaxial strain, and reconstruction were analysed specifically. The Sr-concentration dependence of the critical thickness ($t_c$) of the metal-insulator transition shows a V-shaped behaviour with the smallest $t_c$ = 6 uc at $x$ = 0.5. A detailed analysis of the data indicates that this behaviour originates from a competition between the electronic reconstruction and correlation at the interface. By examining the Hall effect of the LSTO/STO heterostructure, we observed the coexistence of two different types of carriers further confirming the competing relationship between the two mechanisms. Therefore, the optimized electronic properties in the ultrathin oxide system are achieved by modulating the doping concentration and thickness in the LSTO/STO heterostructure.

We first investigated LSTO thin films on $TiO_2$-terminated (001)-orientated STO substrates. The films were grown by pulsed laser deposition (PLD) equipped with *in-situ* reflective high-energy electron diffraction (RHEED). Polycrystalline Sr-doped LTO targets of various stoichiometries were synthesised by mixing high-purity $La_2O_3$ (99.999%), $TiO_2$ (99.99%), and SrO (99.9%) powders. The mixed powders were then sintered at 600 and 900 °C for 8 and 10 hours, respectively, and then pressed into a pellet followed by 36 hours of calcination at 1300 °C. Prior to the deposition, a buffered-HF treatment and air annealing were conducted to obtain pristine $TiO_2$-terminated STO substrates. The growth temperature for all depositions was kept at 850 °C under an oxygen partial pressure of $10^{-4}$ Torr (including during the warming up and cooling down steps). After the growth, the samples' temperature was then lowered to room temperature at a ramping rate of 10 °C/minute. Fig. 1(a) shows a representative RHEED oscillation obtained during the growth process, demonstrating a layer-by-layer growth mode. The thickness of the LSTO was determined by counting the number of the RHEED oscillations. Fig. 1(b) displays an atomic force microscope (AFM) image of the 12 uc LSTO ($x$ = 0.1). Clear atomic terraces are observed, indicating the high structural quality of the LSTO thin films. Transport measurements were conducted using a Quantum Design Physical Property Measurement System (PPMS) in a Van der Pauw configuration. A measurement current of 1 µA was applied over a temperature range from 300 to 2 K. The Hall measurement was carried out under a magnetic field ranging from -8 to 8 T.

Figure 2(a) illustrates the sheet resistance, $R_S$, of the LSTO ($x$ = 0.1)/STO as a function of temperature and film thickness. The $R_S$ generally decreases as the temperature decreases, consistent with a metallic behaviour for relatively large thicknesses. As the film thickness reduces, the $R_S$ gradually increases and exceeds our measurement limit of $10^7$ Ω for a thickness below 10 uc. In contrast to the abrupt MIT with thickness as in the case of the LSTO ($x$ = 0.5)/STO heterostructure[31], the MIT for the *x* = 0.1 sample is more gradual with an intermediate semiconducting state (see the 10 uc case in Fig. 2(a) for an example). This gradual transition is attributed to the variation of the doping concentration, resulting from the strong Coulomb interaction[15,21]. Fig. 2(b) summarizes the *x* = 0.1 conductance as a function of the LSTO thickness at 2 and 300 K. The conductance increases gradually with increasing thickness at 300 K, whereas a sharp jump of more than five-folds is observed at 2 K. It is noteworthy that the MIT occurs around the quantum of conductance (QC), $\frac{2e^2}{h}$, indicating a 2D nature of the carriers in the thin films[33].

To quantitatively study the effect of Sr doping on the interfacial electronic properties, LSTO with varied thicknesses and doping concentrations were grown on STO substrates. Figs. 3(a)-(e) show the temperature-dependent $R_S$ (RT) of LSTO with thicknesses ranging from 8 to above 30 uc and doping concentrations from *x* = 0.1 to 0.9. Interestingly, all the heterostructures experience an MIT but at different $t_c$ corresponding to the different electronic properties. Besides, a Fermi liquid behaviour ($R \propto T^2$)[34] is observed in all concentrations for $t \geq 30$ uc, indicating the conductivity is bulk like. Despite these similarities, one major difference is that the abrupt MIT occurs only at a doping concentration of *x* = ~0.5 and becomes more gradual when the concentration level approaches *x* = 0 and 1. Fig. 3(f) summarizes the doping-dependent $t_c$ at 300 and 2 K. This V-shaped curve shows that the smallest $t_c$ appears at *x* = ~0.5 and gradually increases when concentration increases or decreases toward the two ends, reflecting the nonlinear relationship between the Sr concentration and the electronic property.

We propose the V-shaped curve of $t_c$ is a result of the competition between electronic reconstruction (controlled by polar discontinuity) and electronic correlation (controlled by interfacial strain). Below, we will first discuss the effect of polar discontinuity and then extend the discussion into the interfacial strain. Polar discontinuity is widely accepted as the prevalent mechanism for the formation of 2DEG[35-37]. The carriers of LSTO/STO are a combination of 2D carriers arising mainly from the polar discontinuity as well as the 3D carriers due to the dopant. In this scenario, the LTO (LSTO, *x* = 0) exhibits as a Mott insulator possessing an alternating polarity of +e and -e from $(La^{3+}O^{2-})^+$ and $(Ti^{3+}O^{2-}_2)^-$ planes[38]. When *x* = 1, a non-polar insulator due to the neutral terminated planes $(Sr^{2+}O^{2-})^0$ and $(Ti^{4+}O^{2-}_2)^0$. Therefore, the $Ti^{3+}$, which serves as an indicator of polarity and originates from the electronic reconstruction at the interface, could be modulated via Sr doping. As the Sr concentration decreases from $x = 1$ (equivalent to increasing La doping in STO), the carriers are generated by (i) lowing Sr doping (equivalent increasing La doping), leading to higher bulk carriers, and (ii) the increased polar discontinuity, resulting in an increase in 2D carriers. Therefore, the $t_c$ decreases when Sr doping concentration decreases from $x = 1$ to $x = 0.5$, indicating that the polarity from excess $Ti^{3+}$ is a crucial factor in affecting the electronic reconstruction at the interface and tailoring the electronic properties in the LSTO/STO heterostructure (as shown by the trend indicated by the dotted line in Fig. 3(f)). While the $t_c$ is expects to further decrease at low Sr-doping concentration ($x < 0.5$), $t_c$ value unexpectedly increases, and we attribute this increase to the strain effect.

Strain engineering is an effective strategy to study the electronic correlation in a correlated low-dimensional system[39-41]. LTO is well-known as a strongly-correlated system[42]. Its electronic properties are highly dependent on the strain[15], and a 2% compressive strain is found to induce an MIT behaviour in LTO[43]. The MIT is attributed to the strain-induced changes of crystal-field splitting between the lowest Ti $t_{2g}$ levels[44], which directly affect the electron-electron interaction. Therefore, to clarify the effect of interfacial strain on the electronic property, 30 uc LSTO ($x = 0.1$, 3.909 Å) films[45] were deposited onto other insulating substrates, namely $LaAlO_3$(001) (LAO, 3.791 Å), $(LaAlO_3)_{0.3}(Sr_2AlTiO_6)_{0.7}$(001) (LSAT, 3.868 Å), $NdGaO_3$(110) (NGO, 3.859 Å) and $DyScO_3$ (110) (DSO, 3.944 Å). Fig. 4(a) shows the diverse conducting properties as a function of the biaxial strain in the heterostructures. The LSTO/LAO heterostructure exhibits an insulating behaviour, because of the largest 3.11% compressive strain imposed by LAO onto LSTO. Despite a relatively smaller lattice mismatch in the LSTO/DSO (0.89%), the tensile strain introduced by the DSO turns the interface into insulating as well. Fig. 4(b) shows the $R_S$ at 2 K with the corresponding lattice constant of the substrate. The $R_S$ reaches the highest value with the least compressive strain 0.1% from LSTO/STO heterostructure and drops with increasing compressive strain. It originated from the strain-induced variation of the Ti-3$d$ bandwidth, which could dramatically modulate the strength of electronic correlation. Fig. 4 confirms the electronic properties of LSTO are strongly dependent on the different lattice mismatch from the applied strain, which agrees with the previous work that the tensile strain facilitates the insulating behavior[19].

In analogy, with different concentrations of Sr doping, the lattice mismatch between LSTO and STO could be suppressed from the maximum of 1.66% ($x$

= 0) to 0% (x = 1). The compressive biaxial strain affects the electronic properties of LSTO via changing lattice parameters, resulting in structural relaxations from the corresponding strain. From the band structure perspective, both orbital degeneracy and its occupation are sensitive to the biaxial strain at the interface. Under compressive strain, the two degenerate states in Ti $t_{2g}$ reach a higher energy level so that the crystal-field splitting declines[46]. The energy degeneracy of Ti $t_{2g}$ orbitals results in a small bandwidth, leading to the larger critical thickness. As decreasing the Sr concentration, the compressive strain is intensified. A larger compressive strain facilitates the reduction of bond length between Ti and O and consequently enhances the electron correlation (See Fig. S2 in supplemental Material for details[47])[17,48]. Thus, the continuous promotion of the biaxial compressive strain achieved by decreasing Sr doping at low Sr-doping concentration (x < 0.5) is expected to increase the $t_c$, because the LSTO approaches to the compressively strained LTO in this range. Our observation fits the tendency well at the low Sr-doping concentration range, indicating the correlation of the biaxial strain to the electronic properties (as shown by the trend indicated by the dashed line in Fig. 3(f)).

Taking into account the above mentioned two effects of interfacial strain and polar discontinuity, as the doping concentration increase, the compressive strain results in a smaller $t_c$ while the decreased polarity leads to a larger $t_c$. Consequently, the V-shaped behaviour of the doping-concentration dependent $t_c$ reveals a competition between the electronic correlation and reconstruction coexisting in the LSTO/STO heterostructure.

To further uncover the competition, we examined the carrier density (n) and mobility (μ) by varying the LSTO thicknesses and temperatures. Fig. 5(a) shows the Hall resistance ($R_{xy}$) of LSTO (x = 0.1) films at 10 K with thicknesses ranging from 12 to 300 uc. As the film thickness increases, $R_{xy}$ below 100 K evolves from linear to nonlinear. The nonlinear Hall effect is typically ascribed to multi-channel carriers with different mobilities in this scenario[31]. If two conduction channels are taken into account, the nonlinear $R_{xy}$ (B) can be expressed as[49]:

$$R_{xy} = \frac{(\mu_1^2 n_1 + \mu_2^2 n_2) + (\mu_1 \mu_2 B)^2 (n_1 + n_2)}{e[(\mu_1 |n_1| + \mu_2 |n_2|)^2 + (\mu_1 \mu_2 B)^2 (n_1 + n_2)^2]}$$

, where $n_1$ and $\mu_1$, $n_2$ and $\mu_2$ are the carrier density and mobility of the two types of carriers. Together with the constraint of $R_{xx}$ (0) = $1/e(n_1\mu_1 + n_2\mu_2)$, the $R_{xy}$ (B) data can be fitted. The experimental data in Fig. 5(a) perfectly aligns with the fitting curve, confirming that the nonlinearity indeed arises from two types of carriers coexisting in the LSTO/STO heterostructures. Further information for the carrier density of thin film ($t$ = 12 uc) and bulk form ($t$ = 60 uc) are shown in Figs. 5(b) and 5(c), respectively. While only one type of carrier is observed in the 12 uc LSTO sample, it's clear that two types of carriers coexist in the 60 uc LSTO/STO heterostructure, because of the conducting behaviour in the bulk LSTO. The carrier shows a strong temperature-dependence on both density ($n_1$, $n_2$) and mobilities ($\mu_1$, $\mu_2$), in stark contrast to conventional semiconductor heterostructures[50,51]. For the 60 uc LSTO shown in Fig. 5(c), the density $n_1$ of the majority carriers (solid circle) decreases to $10^{14}$ cm$^{-2}$ upon lowering the temperature, while the mobility $\mu_1$ increases and eventually saturates at ~100 cm$^2$V$^{-1}$s$^{-1}$. In the case of Fermi liquid, where electron-electron scattering is the dominant transport mechanism, electron density decreases as the temperature is lowered. Therefore, the observation of the temperature-dependent majority carriers property, which is a characteristic of a Fermi liquid system, supports the LSTO as a conductor in bulk form and in good accordance with previous reports[52,53]. Thus, we conclude that the majority of carriers are 3D-type and exist in the LSTO bulk. The minority carrier ($n_2$) starts to increase as the temperature drops below 100 K and saturates at ~5 × $10^{12}$ cm$^{-2}$, more than 10 times lower than $n_1$. Meanwhile, $\mu_2$ of the minority carrier only increases slightly with decreasing temperature and seems independent of $n_2$. These results suggest that the minority carrier is a 2D electron gas rather than a 3D Fermi liquid.

Figure 5(d) shows the thickness-dependent $n_1$ and $n_2$ of LSTO ($x$ = 0.1)/STO heterostructure collected at 10 K. As the film thickness increases, the effect of biaxial strain weakens, resulting in the increase of 3D carrier density ($n_1$, red circle) and a decrease in mobility ($\mu_1$, black circle). Conversely, the density ($n_2$, red triangle) and mobility ($\mu_2$, black triangle) of the 2D carrier are enhanced with the increasing thickness of LSTO films. Fig. 5(e) shows the $R_{xy}$ of 30 uc bulk LSTO films with Sr-doping concentrations ranging from $x$ = 0.1 to 0.9 (*i.e.* 10%, 50%, 75% and 90% of Sr) at 2 K. A clear nonlinear Hall effect is observed in all concentrations as well, confirming the coexistence of 2D and 3D carrier in LSTO/STO heterostructure. Fig. 5(f) further elucidates the property of two types of carriers with corresponding Sr doping levels. In the low concentration region ($x \leq 0.5$), the 3D carrier density slightly increases and reaches the peak at 50%, while the 2D carrier density decreases as the doping concentration increases, indicating a suppression of polarity with increasing Sr doping. In the high concentration region ($x > 0.5$), both the two types of carriers decline with the increasing doping concentration. It indicates the strain effect plays a more important role in the low concentration regime while with a higher doping concentration, the interfacial polar discontinuity plays a more significant role on the electronic properties of LSTO/STO heterostructure, which is consistent with the observed V-shaped behaviour of $t_c$.

# Conclusion

In summary, by varying the Sr-doping concentration in the LSTO/STO heterostructures, we demonstrate that competition between electronic correlation (modulated by strain) and reconstruction (modulated by polarity) is responsible for the V-shaped behaviour of the $t_c$. Specifically, as the doping concentration drops, the strain effect raises the $t_c$ of the MIT, while the interfacial polar discontinuity has the opposite effect. Moreover, we found that the two types of electrons coexist in the LSTO/STO heterostructures, namely a 3D Fermi liquid in the bulk LSTO and a 2DEG at the interface. Our study paves the way to further create and utilise the unique electronic properties in complex oxide heterostructures by leveraging the interplay between electronic reconstruction and electronic correlation.

# Figures

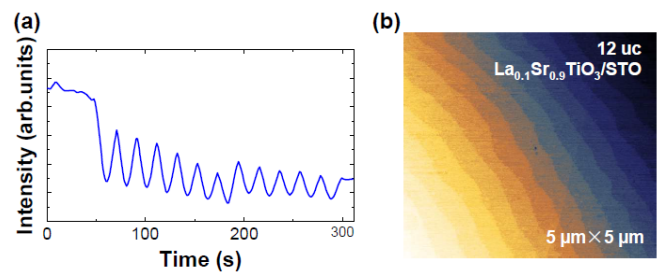

FIG. 1. Reflection high-energy electron diffraction (RHEED) oscillations and surface morphology of the 12 uc $La_{1-x}Sr_xTiO_3$ (LSTO, $x$ = 0.1)/STO heterostructure. (a) RHEED oscillations of the 12 uc LSTO ($x$ = 0.1) grown on STO substrate. (b) The atomic force microscopy (AFM) image of 12 uc LSTO ($x$ = 0.1)/STO.

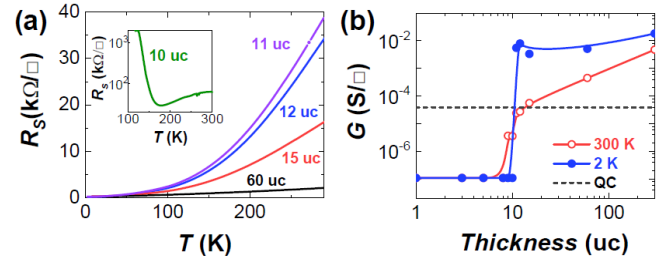

FIG. 2. Metal insulator transition (MIT) of LSTO (x = 0.1)/STO heterostructure. (a) Temperature-dependent $R_S$ (RT) curves of samples with a critical thickness ($t_c$) of MIT at 10 uc. Samples with LSTO thickness below 10 uc show an insulating behaviour, and above 10 uc a metallic behaviour. (b) $R_S$ at 300 K and 2 K as a function of LSTO thickness. An abrupt MIT transition is observed at 2 K. The quantum of conductance (QC), $\frac{2e^2}{h}$, is plotted as a guide to the eyes (black dashed line).

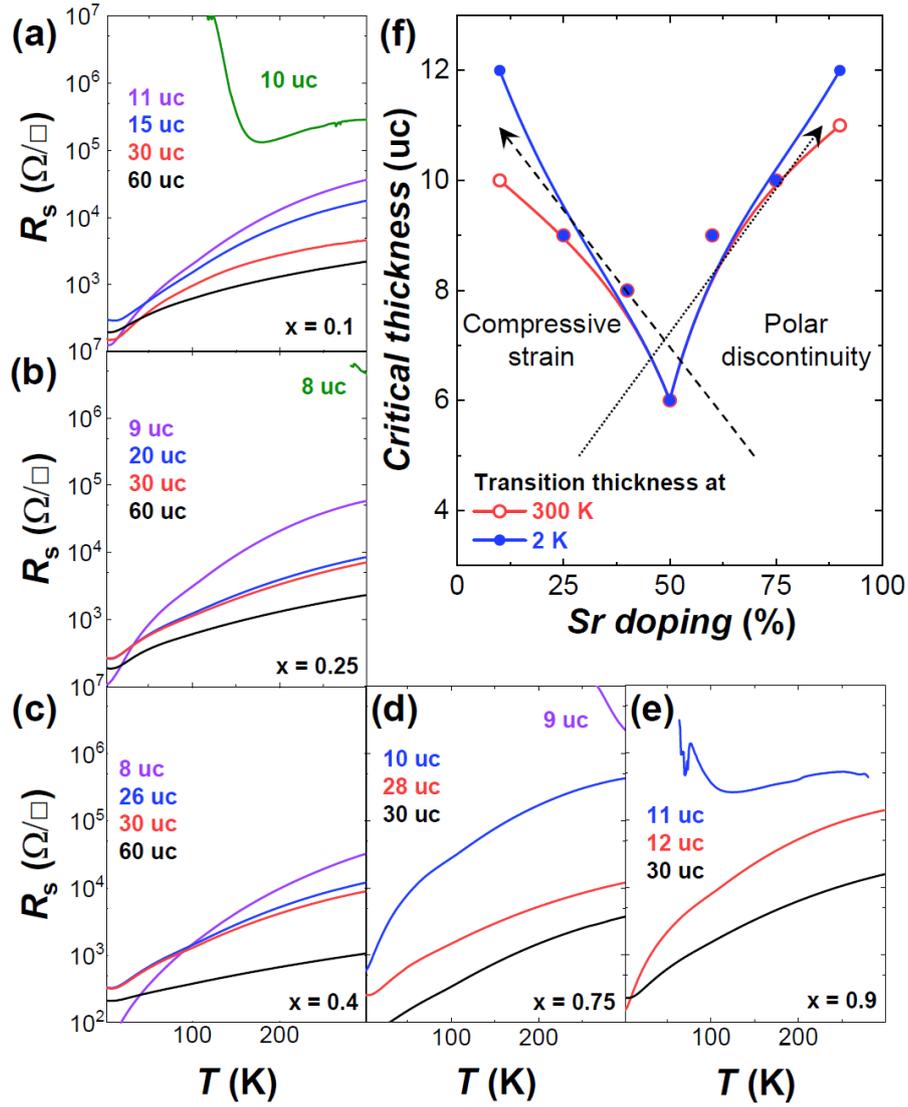

FIG 3. RT of LSTO with different thicknesses and doping concentrations. (a)-(e) Temperature-dependent $R_S$ from ultrathin to bulk thickness with Sr-doping concentrations from 10 to 90%, namely 10% ($x = 0.1$), 25% ($x = 0.25$), 40% ($x = 0.4$), 50% ($x = 0.5$), 75% ($x = 0.75$) and 90% ($x = 0.9$). (f) Summary of $t_c$ as a function of doping concentration at both 300 and 2 K.

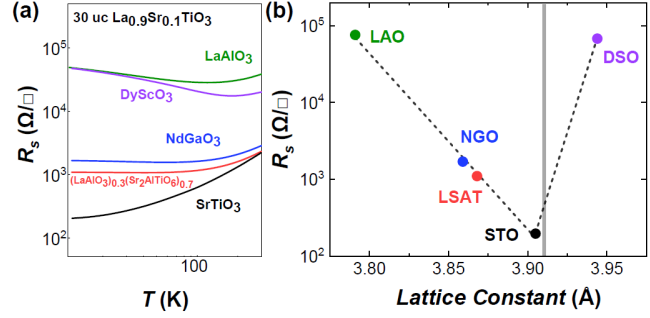

FIG. 4. The effect of strain on the electrical transport properties of the bulk from (t = 30 uc) of LSTO (x = 0.1). (a) RT curves of 30 uc LSTO (x = 0.1) grown on different substrates. There is a clear rise of resistance as the strain increases. (b) The lattice mismatch of 30 uc LSTO (x = 0.1) deposited onto different substrates, namely LAO, NGO, LSAT, STO and DSO.

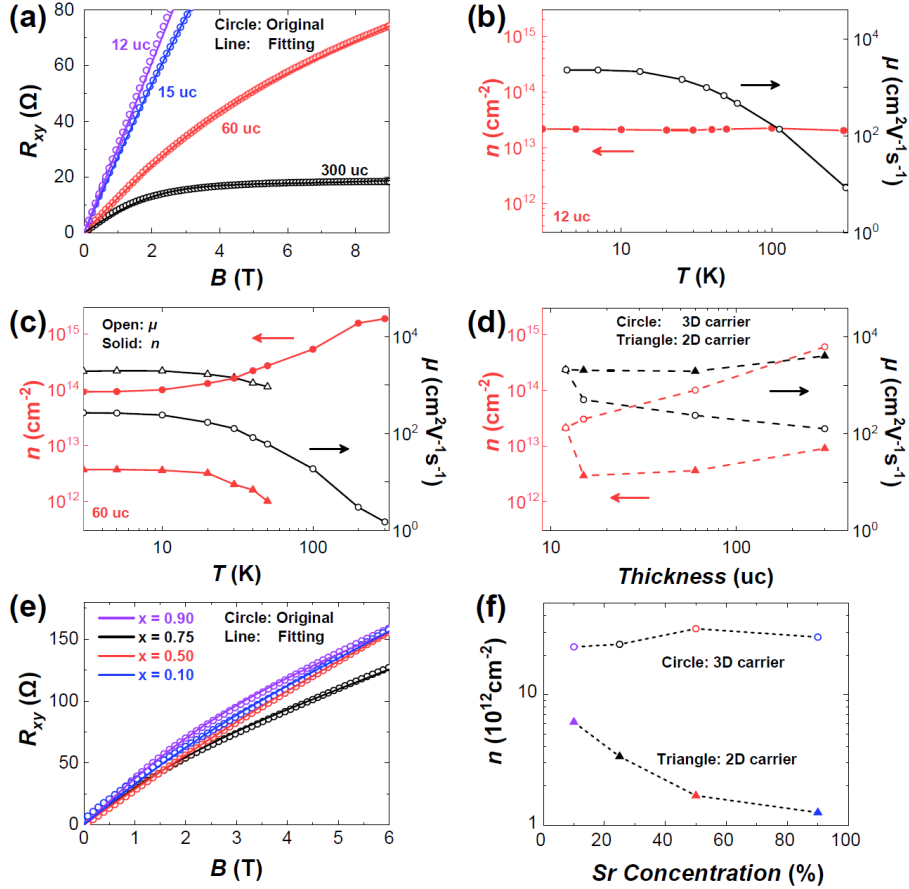

FIG. 5. Carrier density ($n$) and mobility ($\mu$) of LSTO as functions of thickness and temperature. (a) $R_{xy}(B)$ as a function of the magnetic field for different film thicknesses measured at low temperatures. All samples show nonlinear behaviour. Temperature dependence of $n$ and $\mu$ of 12 uc (b) and 60 uc (c) samples. (d) Thickness dependence of the $n$ and $\mu$ of the two types of carrier

measured at low temperatures. (e) $R_{xy}$ of bulk LSTO (t = 30 uc) for various concentrations, *i.e.* 10%, 50%, 75% and 90%, at 2 K. All samples show nonlinear behaviour. (f) *n* of two types of the carrier as a function of Sr-doping concentration.

**Acknowledgements**

**Funding:** X.R.W. acknowledges support from the Nanyang Assistant Professorship grant from Nanyang Technological University, Academic Research Fund Tier 1 (RG108/17 and RG177/18) from Singapore Ministry of Education, Singapore Ministry of Education under its Academic Research Fund (AcRF) Tier 2 (Grant Nos. MOE-T2EP50120-0006 and MOE-T2EP50220-0005) and Tier 3 (Grant No. MOE2018-T3-1-002), and the Agency for Science, Technology and Research (A*STAR) under its AME IRG grant (Project No. A20E5c0094).

**Author contributions:** A.A. X.R.W and X.Y.W conceived and designed the experiments. C.Y., L.S. and Z.H. performed the electrical transport property measurement. X.Y.W. and X.R.W. wrote the manuscript with help from all other authors. All authors contributed to the discussion and interpretation of the results.

**Competing interests:** The authors declare that they have no competing interests.